\documentclass{article} 
\usepackage{iclr2024_conference,times}

\usepackage{amsmath,amsfonts,bm}

\def\eqref#1{equation~\ref{#1}}

\def\1{\bm{1}}

\DeclareMathAlphabet{\mathsfit}{\encodingdefault}{\sfdefault}{m}{sl}
\SetMathAlphabet{\mathsfit}{bold}{\encodingdefault}{\sfdefault}{bx}{n}

\usepackage{hyperref}
\usepackage{url}
\usepackage{booktabs}
\usepackage{wrapfig, multirow}
\usepackage{pifont}
\usepackage{array}
\newcommand{\xmark}{\ding{55} \space}

\title{Identifying Climate Targets in National Laws and Policies using Machine Learning}

\author{Matyas Juhasz\thanks{\texttt{dsci@climatepolicyradar.org}} \\
Climate Policy Radar \\
\And{}Tina Marchand\thanks{Was affiliated with Climate Policy Radar when this work was carried out.} \\
\And{}Roshan Melwani \\
Climate Policy Radar \\
\AND{}Kalyan Dutia \\
Climate Policy Radar \\
\And{}Sarah Goodenough \\
Climate Policy Radar \\
\And{}Harrison Pim \\
Climate Policy Radar \\
\And{}Henry Franks \\
Climate Policy Radar \\
}

\iclrfinalcopy 
\begin{document}

\maketitle

\begin{abstract}
Quantified policy targets are a fundamental element of climate policy, typically characterised by domain-specific and technical language. Current methods for curating comprehensive views of global climate policy targets entail significant manual effort. At present there are few scalable methods for extracting climate targets from national laws or policies, which limits policymakers' and researchers' ability to (1) assess private and public sector alignment with global goals and (2) inform policy decisions. In this paper we present an approach for extracting mentions of climate targets from national laws and policies. We create an expert-annotated dataset identifying three categories of target ('Net Zero', 'Reduction' and 'Other' (e.g. renewable energy targets)) and train a classifier to reliably identify them in text. We investigate bias and equity impacts related to our model and identify specific years and country names as problematic features. Finally, we investigate the characteristics of the dataset produced by running this classifier on the Climate Policy Radar (CPR) dataset of global national climate laws and policies and UNFCCC submissions, highlighting the potential of automated and scalable data collection for existing climate policy databases and supporting further research. Our work represents a significant upgrade in the accessibility of these key climate policy elements for policymakers and researchers. We publish our model and related dataset \footnote{Model: \url{https://huggingface.co/ClimatePolicyRadar/national-climate-targets}; dataset: \url{https://huggingface.co/datasets/ClimatePolicyRadar/national-climate-targets}}.
\end{abstract}

\section{Introduction}

Climate law and policy are a primary lever for national and international climate action. Effective laws and policies are typically formulated with reference to (1) what has already been implemented across different jurisdictions, and (2) policy adequacy and effectiveness, including in comparison to global trends and international frameworks such as the Paris Agreement. Targets – quantified, measurable expressions of prospective policy outcomes – are a cornerstone of effective climate policy. Targets bolster the credibility of countries’ commitments by setting quantifiable objectives, and inform policy design, implementation and monitoring (\cite{nachmany2018}, \cite{anderson2021}, \cite{haarstad2020}). This is true of economy-wide Greenhouse Gas (GHG) targets, such as those in Nationally Determined Contributions (NDCs) submitted under the UN Framework Convention on Climate Change (UNFCCC), and of sector-specific targets such as in transportation or agriculture. Data on existing targets addressing climate change is invaluable for policy analysis, and actors engaged in the design or evaluation of national laws and policies often look to targets as a first ‘port of call’ to establish national and international progress and ambition in addressing climate change. There follows a need for effective tools to support the creation of climate target datasets.

Analysis of existing laws and policies is constrained by the (often limited) resources and expertise at the disposal of relevant actors. Common barriers include: (1) key information buried within lengthy, difficult-to-parse documents, and (2) policies and targets being written in many different languages and terminologies (e.g. "reduce emissions by 50\% by 2030, against a 2005 baseline" versus "double energy production from renewable sources in the next 5 years"). These constraints on time and resources affect all actors working to understand and improve the efficacy of climate laws and policies, including policymakers, academics, NGOs and UN bodies, and are especially acute for those operating under resource constraints, including in low-income countries and communities.

In this paper we present an approach for extracting mentions of targets from national climate laws and policies, using paragraph-level classification. We publish our model and related dataset \footnote{\footnote{Model: \url{https://huggingface.co/ClimatePolicyRadar/national-climate-targets}; dataset: \url{https://huggingface.co/datasets/ClimatePolicyRadar/national-climate-targets}}}.

\section{Related work}
Existing databases, including Net Zero Tracker (\cite{Lang2023NetZeroTracker}), Climate Change Laws of the World (CCLW, \cite{cclw2023}), ClimateWatch (\cite{ClimateWatch2022}), and Climate Action Tracker (\cite{climateactiontracker}) rely on manual extraction of targets (in some cases relying on volunteers), limiting their scalability and consuming significant organisational resources. As a result, many only collect targets from NDCs and focus only on economy-wide targets (and not on other types of targets like sectoral targets). This results in a fundamental gap in the analysis of global public and private sector commitments. 

Machine Learning (ML) systems are an attractive solution supporting the automated extraction of structured information at scale and facilitating a more effective, exhaustive, and rigorous analysis than what is attainable via the application of human labour. Natural Language Processing (NLP) has recently been applied successfully in a variety of research contexts including climate and policy (e.g. \cite{cody-climate-sentiment}, \cite{sietsma2023gst}, \cite{bingler2022cheaptalk}), rendering analysis of the global policy landscape tractable (\cite{biesbroek2020ukadaptationpolicy}). NLP is especially applicable here as stakeholders must filter, analyse, and manage vast amounts of unstructured text (\cite{stede2021climate}). The advent of transformers (\cite{vaswanitransformers}), which substantially improve the power of models in natural language understanding, has opened up significant new avenues for the application of NLP to the climate policy domain: identifying nuanced language features in large scale (n  \textgreater{} 100,000) document datasets (\cite{callaghan2021machine}), quantifying regulatory climate risk disclosures (\cite{kolbel2020ask}, or analysing sustainability reports (\cite{luccioni2020analyzing}). Text classification has also been used to facilitate analysis of government and corporate climate policy (\cite{kolbel2020ask}, \cite{stammbach2022claims}, \cite{spokoyny2023climabench}). 

NLP has been applied to extract climate-related targets before (\cite{schimanski2023climatebertnetzero}), but without addressing individual GHGs or economic sectors. Our work extends existing contributions by (1) extending the definitions of emissions reduction and net-zero targets to include those also addressing individual GHGs and sectors of the economy, as these more specific targets are an important component of a country's ambition and ability to achieve its economy-wide emissions reduction targets (\cite{ipcc2023}); (2) introducing a new 'Other' category of targets to capture quantified targets made by governments with mitigation or adaptation objectives that do not explicitly mention emissions reduction, such as reducing deforestation or scaling up renewable energy capacity; and (3) curating a multi-label dataset, enabling each instance to be associated with zero, one, or multiple designated categories.

\section{Data}

The data used for this project was sourced from the Climate Policy Radar (CPR) database \cite{cpr2024} of national laws and policies and UNFCCC submissions containing over 4,000 documents published by every single national government. We assign the target types "Net Zero", "Reduction" or "Other" to paragraphs in a multi-label classification setting. A target satisfies three criteria: it (1) contains an aim to achieve a specific outcome, (2) is quantifiable, and (3) has been given a deadline. We consider targets set by governments focusing on their specific national objectives and actions, rather than referring to a collective regional or global goal. Reduction targets refer to a reduction of GHG emissions. They can be economy-wide or sector-specific, and refer to different types of GHGs. Net Zero targets constitute a commitment to balance GHG emissions with removal, effectively reducing the net emissions to zero. "Other" targets are those that do not fit into the Reduction or Net Zero category, yet satisfy the three criteria, for example renewable energy targets. See Appendix \ref{appendix-policy-methodology} for detailed definitions.

The low relative frequency of targets renders sampling for annotation a challenge. Our approach took CCLW's manually written target summaries (\cite{cclw2023}) as seeds, matching them to paragraphs in the corresponding documents. Stratified sampling was employed to oversample for underrepresented regions and machine translated documents. Negative sampling was used to adjust the label ratio. Three domain-expert annotators labelled a 2,610 paragraphs containing 1,193 targets (Table \ref{dataset-table}) using the following process: after initial labelling, a subset of the dataset labels underwent a review, and inter-annotator agreement was measured throughout the process to ensure consistency. Subsequently, we employed rounds of active learning to sample and annotate additional examples.

\begin{table}[t]
\begin{center}
\begin{tabular}{lllll}
\toprule
\multicolumn{1}{c}{\bf Net Zero} &\multicolumn{1}{c}{\bf Reduction} &\multicolumn{1}{c}{\bf Other} &\multicolumn{1}{c}{\bf No annotation} &\multicolumn{1}{c}{\bf Total paragraphs} \\
\midrule
 203 &        359 &    631 &       1584 &   2610 \\
\bottomrule
\end{tabular}
\caption{Counts of labels in the final dataset. 'No annotation' counts paragraphs where no target was found.}
\label{dataset-table}
\end{center}
\end{table}

\section{Classifier Training}

\begin{wraptable}{r}{0.5\textwidth}
\centering
\renewcommand{\arraystretch}{1.25}
\begin{tabular}{>{\bfseries}llr}
\toprule
\bf{Class} & metric & value ($\sigma$) \\
\midrule
\multirow{3}{*}{NZT} & precision &  0.777 (0.043) \\
    & recall &  0.911 (0.042) \\
    & f1 &  0.837 (0.023) \\
\cline{1-3}
\multirow{3}{*}{Reduction} & precision &  0.827 (0.063) \\
    & recall &  0.914 (0.045) \\
    & f1 &  0.867 (0.036) \\
\cline{1-3}
\multirow{3}{*}{Other} & precision &  0.801 (0.022) \\
    & recall &  0.889 (0.016) \\
    & f1 &  0.842 (0.008) \\
\cline{1-3}
\multirow{3}{*}{all} & precision &   0.803 (0.020) \\
    & recall &    0.900 (0.003) \\
    & f1 &  0.849 (0.012) \\
\bottomrule
\end{tabular}
\caption{Classifier performance on the three annotation classes: net zero target, reduction target and other target.}
\label{class-performance-table}
\end{wraptable}

To accommodate a significant overlap across labels, we used a multi-label text-classification approach. We ran a number of experiments to select the most appropriate base model for our text classification task. We report results of our comparison in Appendix \ref{appendix-model-comparison}. Selecting climateBERT (\cite{wkbl2022climatebert}), we ran a grid-search to identify the optimal hyperparameters for fine-tuning, set out in Appendix \ref{appendix-model-hyperparameters}.

Our model effectively predicts all 3 labels with an overall f1 score of 0.849 (Table \ref{class-performance-table}). The lower performance of the NZT label is due to the low prevalence of such targets in climate text, entailing a low volume of training data. This category is particularly prone to the model learning erroneous relationships, as discussed in Section \ref{impacts-and-equity}.

\section{Impacts \& Equity Considerations}
\label{impacts-and-equity}

\textbf{Biases toward countries and round years.}
Despite the stratified sampling, targets are less prevalent for documents authored by countries in the global south. Models trained on the dataset subsequently attribute higher probabilities to a paragraph containing targets referencing specific country names.

Targets frequently reference years that are multiples of 5 (e.g. 2035 or 2050), and models trained on the dataset can learn these features as predictors. We hypothesise a bias towards round dates in the pretrained RoBERTa model: when predicting masked years, both distilRoBERTa and climateBERT consistently predicted round years more confidently than non-round years. Our analysis shows that "2020", "2021", "2030" and "2050" (and no others between 2020 and 2100) are single tokens in distilRoBERTa's vocabulary, potentially biasing model behaviour. 

\textbf{The effect of machine translation.} Our dataset contains English paragraphs, sourced from English documents (65.29\%) and Google Cloud Translation API (\cite{googlecloudtranslate}) translated documents (34.71\%) from 37 source languages. While there is a drop in Overall F1 score associated with classifying machine translated text (Table \ref{translation-performance-table}), this is small (0.023). Labels are disproportionately impacted with a bigger drop in "Net Zero" (0.078), medium in "Reduction" (0.041) and insignificant (0.001) in "Other". It would be valuable to investigate whether balancing translated text and original language in the training data could address the observed drops in performance.
 
\begin{table}[h]
\begin{center}
\begin{tabular}{lrrrrrrrr}
\toprule
{} & \multicolumn{2}{c}{\bf Net Zero} & \multicolumn{2}{c}{\bf Reduction} & \multicolumn{2}{c}{\bf Other} & \multicolumn{2}{c}{\bf Overall} \\
{} & \textbf{count} &     \textbf{F1} &     \textbf{count} &     \textbf{F1} & \textbf{count} &     \textbf{F1} & \textbf{count} &     \textbf{F1} \\
\midrule
Original language &   153 &  0.856 &       257 &  0.880 &   401 &  0.843 &   811 &  0.857 \\
English  &    50 &  0.778 &       102 &  0.839 &   230 &  0.842 &   382 &  0.834 \\
\bottomrule
\end{tabular}
\caption{Classifier performance on original language vs machine-translated text. Column `count' is the number of test samples for each class and in total.}
\label{translation-performance-table}
\end{center}
\end{table}
\section{Model Application}

To investigate potential applications of this model we produced a dataset by running the model across all the text of all the CPR laws, policies and UNFCCC submissions the labelled data was sampled from. We did so at a threshold which mapped to 80\% precision and 80\% recall, as evaluated against the validation set. The resulting dataset contained 24,583 target mentions published by 201 national governments, broken down into 5,223 net zero, 7,019 reduction and 13,617 'Other' types.

We further applied topic modelling using BERTopic (\cite{grootendorst2022bertopic}, see Appendix \ref{appendix-metadata-extraction}) over the paragraphs with target type 'Other'. The most common topics we found relate to specific systems and sectors of the economy: Renewables (general) (31.4\% of 'Other' target mentions); Agriculture, Forests \& Fisheries (13.4\%); Miscellaneous (12.1\%); Transport (11.2\%); and Electricity, Infrastructure and Energy Efficiency (11.0\%). Using automatically extracted targets, practitioners can therefore identify and compare trends, gaps and opportunities for national climate action between specific systems or sectors.

\section{Conclusion}

In this paper we present an approach for extracting targets from climate law and policy documents, by identifying Net Zero, Reduction, and/or Other types of target. We further probe our dataset and model for underlying bias and equity concerns. Supporting scalable analysis of climate law, policy and UN submissions is key to helping governments enact effective laws and policies and to support the evaluation and accountability of global climate action. Our approach to classifying targets within law, policy and UN submission documents allows for rapid identification of target language in large document corpora, which can play a role in improving collective understanding of the gaps and opportunities for enhancing national ambition and efficacy in addressing climate change.

Our model enables much more detailed analysis of climate targets than was previously possible, through extending previous work to include targets related to specific GHGs, sectors, and a large number of non-GHG-related targets. We analyse a dataset produced by running this model across a large dataset of climate laws and policies, demonstrating its utility to policymakers and researchers to assess areas such as targets specific to high-emitting sectors and technologies, such as agriculture and transport.

This work enables automated data collection for existing, manually curated climate databases, as well as further research applying machine learning to aid climate policy analysis. An important avenue of climate policy analysis that this work enables is identifying discrepancies between the ambition of targets set by national governments in their submissions to the UN Climate Change Secretariat (most commonly in their Nationally Determined Contributions (NDCs)), and of the targets in their national laws and policies. This "implementation gap" is an important (\cite{fransen_taking_2023}) but previously time consuming and manual research challenge. Other future work includes (1) predicting targets made by other actors such as companies and cities, states and regions, (2) further NLP analysis of large, machine-produced datasets such as the one presented in this work, and (3) extracting structured representations of targets for additional analysis, such as extracting target deadlines or segmenting by specific GHG references. 

\bibliography{iclr2024_conference}
\bibliographystyle{iclr2024_conference}

\appendix
\section{Appendix}

\subsection{Training Data Collection}
\label{appendix-training-data-collection}

\subsection{Methodology Used for Labelling}
\label{appendix-policy-methodology}

The definitions used for labelling are based on existing work by Net Zero Tracker (\cite{Lang2023NetZeroTracker}) and ClimateBERT-NetZero (\cite{schimanski2023climatebertnetzero}) to identify net zero and emissions reduction targets. We extend these definitions to include a new class – ‘Other’ – to capture all other quantified targets made by national governments in climate policy documents.

We also expand the net zero and emissions reduction targets definitions to include targets for different greenhouse gases (as well as general greenhouse gas targets) and to include sector-specific targets (such as emissions reduction targets for the transport sector).

\subsubsection{Definition of a target}

\begin{description}

\item[A target is] \hfill \\

\textbf{An aim to achieve something}, rather than stating something concrete about the future. Often this means that the phrase indicates a level of uncertainty.

Examples 

\begin{itemize}
    \item  \checkmark \space Food waste reduced by 10\% by 2022 and another 20\% by 2030
    \item  \checkmark \space Not less than 25,000 new jobs created in 5 years
    \item  \checkmark \space The Government will endeavour to reach a minimum level of 10\% of electrical energy supplied to the grid to be from NRE by a process of facilitation including access to green funding such as CDM.
    \item  \xmark  It is anticipated that industrial production will increase by a minimum 4.6\% annually.
    \item \xmark  Life expectancy of our people by 2040 will be 80 years due to quality care for older generation, a decent level of pension benefits and a high degree of family care.
    \item \xmark  The Startup \& CSI Development Flagship Programme is expected to create about 4,700 additional jobs in existing CSIs, Startups and new CSIs within the 12th FYP period.
    \item \xmark  It is anticipated that industrial production will increase by a minimum 4.6\% annually.
\end{itemize}

\textbf{Quantifiable}: it contains a reference to a measurable quantitative value. This may be numeric or non-numeric. For example, words like all, every, double, halve, eradicate, no, none and independent of refer to measurable quantities. 

Examples 

\begin{itemize}
    \item  \checkmark \space reduce emissions by 68\% by 2030
    \item  \checkmark \space provide piped water supply to all rural households by 2024
    \item \xmark  Credit Guarantee Enhancement Corporation to be set up in 2019-2020.
    \item \xmark  significantly decrease food waste to reduce emissions by the next decade
\end{itemize}

\textbf{Given a deadline}: it aims to achieve something quantifiable by a certain point in time. It can be expressed through a specific end date or some other representation of an end date in reference to planning cycles or number of years.

\begin{itemize}

    \item \checkmark \space reduce emissions by 68\% by 2030
    \item \checkmark \space in the next ten years, we will add 100km of new bicycle lanes
    \item \checkmark \space increase renewable energy capacity by 20\% by the end of the current national 5 year planning cycle
    \item \xmark increase amount of protected areas by a minimum of 4.6\%
    \item \xmark reach energy efficiency savings of at least 2\% on an annual basis

\end{itemize}

\item[A target is not] \hfill \\
\begin{itemize}

\item A policy action or a commitment to perform one (e.g. \textit{"Publish the government’s low carbon transition plan for the period 2020-2025."}).
\item An abstract reference with no information about what the target is (e.g. \textit{"Montenegro’s compliance schedule will run parallel to that of EU members in the 2020-2030 decade so as to, jointly, reach the international targets established for 2030."}).
\item An analysis of a target (e.g. \textit{"It can be seen from Figure 10-3 that while the average RE cost of the MEPU 40\% target is higher than the average RE cost of the MEPU 35\% target, the average system cost for the 40\% target is only marginally higher than the 35\% target."}).
\item A commitment to set up a vague target in the future (e.g. \textit{"This assumes that the tighter EU ETS cap agreed as part of an EU deal on moving to a 30\% target would continue at the same rate of reduction beyond 2020."}).
\item A commitment to achieve a target based on the fulfilment of certain conditions (e.g. \textit{"if we receive international finance, then we would be ready to achieve further GHG emissions reduction of 35\% by 2040, compared to 2005 levels."}).

\end{itemize}

\end{description}

\subsubsection{Target types}

\begin{description}
  \item[Reduction] \hfill \\ An emissions reduction target is a claim made by a public institution that refers to a reduction in GHG emissions by a certain point in time. It can be expressed as an absolute or relative reduction of GHG emissions, sometimes benchmarked against a baseline year or a business as usual (BAU) scenario to which the reduction target is compared. It can also be expressed as an emissions intensity reduction target where emissions act as the numerator and something else (typically population, GDP, or revenue) as the denominator. The emissions reduction target can be economy-wide or sectoral, and it can also refer to different types of GHGs (e.g. carbon dioxide, CO\textsubscript{2}eq, methane). Must be a national target, not global.
  \item[Net zero] \hfill \\ A net-zero target is a special type of emissions reduction target where a public institution states to bring its emissions balance down to no additional net emissions by a certain year. The net-zero target can be economy-wide or sectoral. We take particular care with mentions of net-zero technologies, they are not net-zero sectoral targets. Must be a national target, not global. To be considered a net-zero target, the emissions reduction target must contain reference to this scoped language:
  \begin{itemize}
      \item Net zero
      \item Carbon neutral(ity)
      \item GHG neutral(ity)
      \item Greenhouse gas neutral(ity)
      \item Carbon negative
      \item Net negative
      \item Carbon free
      \item Zero (or 0) emissions
      \item Zero (or 0) carbon
      \item Fully decarbonise
      \item Climate neutral
      \item Climate positive
      \item 100\% emissions reduction
  \end{itemize}
  \item[Other] \hfill \\ Refers to cases where a public institution aims to achieve something concrete that is both quantifiable and has a deadline. This could include, but not limited to, non-climate mitigation (emissions reduction or net zero) targets, such as an adaptation, nature-based or renewable energy target. It could also include a policy measure, such as a quantifiable increase in carbon pricing by a certain time as a way to support the achievement of an overall emissions reduction target. Must be a national target, not global.
\end{description}

\subsection{Model performance comparisons}
The models investigated were RoBERTa-base (\cite{DBLP:journals/corr/abs-1907-11692}),  DistilRoBERTa-base (\cite{Sanh2019DistilBERTAD}) and climatebert/distilroberta-base-climate-f (\cite{wkbl2022climatebert}), with the model performances summarised in Table \ref{model-comparison-table}. RoBERTa-base had the best performance, only outperforming ClimateBERT by 0.006 on the overall F1 score. Climatebert's size (more than 4x smaller model) and balanced performance (outperforming RoBERTa-base on the "Net Zero" label by 0.029) were the main factors behind our selection of it as the base model.

\label{appendix-model-comparison}

\begin{table}
\begin{center}
\begin{tabular}{lllll}
\toprule
    &    &    ClimateBERT (82.4M) & DistilRoBERTa-base (82.8M) &   RoBERTa-base (355M) \\
\midrule
\multirow{3}{*}{NZT} & precision &  0.777 (0.043) &      0.758 (0.047) &   0.819 (0.02) \\
    & recall &  0.911 (0.042) &      0.754 (0.058) &  0.799 (0.048) \\
    & f1 &  0.837 (0.023) &      0.754 (0.017) &  0.808 (0.021) \\
\cline{1-5}
\multirow{3}{*}{Reduction} & precision &  0.827 (0.063) &       0.81 (0.013) &  0.843 (0.031) \\
    & recall &  0.914 (0.045) &        0.9 (0.038) &  0.911 (0.022) \\
    & f1 &  0.867 (0.036) &      0.852 (0.014) &  0.876 (0.027) \\
\cline{1-5}
\multirow{3}{*}{Other} & precision &  0.801 (0.022) &      0.807 (0.019) &  0.824 (0.019) \\
    & recall &  0.889 (0.016) &      0.864 (0.047) &   0.895 (0.02) \\
    & f1 &  0.842 (0.008) &      0.834 (0.024) &  0.858 (0.004) \\
\cline{1-5}
\multirow{3}{*}{all} & precision &   0.803 (0.02) &      0.799 (0.012) &  0.829 (0.013) \\
    & recall &    0.9 (0.003) &      0.856 (0.033) &  0.884 (0.021) \\
    & f1 &  0.849 (0.012) &      0.826 (0.014) &  0.855 (0.012) \\
\bottomrule
\end{tabular}
\end{center}
\caption{Model performances per label and overall}
\label{model-comparison-table}
\end{table}

\subsection{Hyperparameters for Model Training}
\label{appendix-model-hyperparameters}

\begin{tabular}{ll}
\toprule
{} &            value \\
\midrule
seed                         &           42 \\
optim                        &  adamw\_torch \\
adam\_beta1                   &          0.9 \\
adam\_beta2                   &        0.999 \\
model\_type                   &      roberta \\
adam\_epsilon                 &          0.0 \\
warmup\_steps                 &          100 \\
weight\_decay                 &         0.01 \\
learning\_rate                &      0.00002 \\
num\_train\_epochs             &            5 \\
lr\_scheduler\_type            &       linear \\
hidden\_dropout\_prob          &          0.1 \\
per\_device\_eval\_batch\_size   &           24 \\
gradient\_accumulation\_steps  &            1 \\
per\_device\_train\_batch\_size  &           16 \\
attention\_probs\_dropout\_prob &          0.1 \\
\bottomrule
\end{tabular}

\subsection{Topic Modelling on 'Other' Targets}
\label{appendix-metadata-extraction}

\subsubsection{Pre-Processing Step}
To ensure processes heuristics were applied to extract sentence likely to contain a quantified target from each paragraph predicted as containing a target. These heuristics were whether the sentence contained the word 'target' (not case-sensitive), or any entity expressing a date, amount or measurement (\texttt{DATE}, \texttt{CARDINAL}, \texttt{QUANTITY}, \texttt{PERCENT}) as predicted by spaCy's (\cite{ines_montani_2023_10009823}) \texttt{en\_core\_web\_lg} model. Paragraphs that metadata extraction were run on were the sentences that the heuristics predicted as containing quantified targets concatenated, or the entire paragraph if no sentences in the original paragraph were predicted by the heuristics.

\subsubsection{Topic modelling}
BERTopic (\cite{grootendorst2022bertopic}, parameters in Table \ref{bertopic-parameters}) was run on pre-processed paragraphs predicted as 'Other'  to generate 60 topics. Seed topics were iteratively refined, and the final list of topics was grouped into 9 higher-level topics, with irrelevant-seeming topics discarded.

These 9 topics, in descending order of frequency, are:
\begin{itemize}
    \item Renewables (general)
    \item Agriculture, forests \& fisheries
    \item Miscellaneous
    \item Transport
    \item     Electricity, infrastructure \& energy efficiency
    \item Waste, water \& plastic
    \item Social wellbeing (health, education and social housing)
    \item Wind \& solar
    \item Built environment \& construction
\end{itemize}

\begin{table}[h]
    \centering
    \caption{BERTopic configuration}
    \label{bertopic-parameters}
    \begin{tabular}{|c|p{6cm}|}
        \hline
        \textbf{Parameter} & \textbf{Value} \\
        \hline
        nr\_topics & 60 \\
        top\_n\_words & 8 \\
        seed topics & 
        \begin{tabular}[t]{@{}l@{}}
            "energy efficiency" \\
            "renewable energy" \\
            "energy sources" \\
            "land use" \\
            "forests", "forest cover" \\
            "deforestation and reforestation"
        \end{tabular} \\
        \hline
        embedding model & sentence-transformers/all-MiniLM-L6-v2 \\
        representation model & KeyBERT (\cite{grootendorst2020keybert}) \\
        vectorizer model & CountVectorizer \\
        vectorizer ngram\_range & (1,3) \\
        vectorizer min\_df & 5 \\
        vectorizer stop\_words & "english" \\
        \hline
    \end{tabular}
\end{table}

\end{document}